\documentclass[10pt]{article}

\input{tcilatex}
\begin{document}

\begin{center}
{\LARGE Quantum interview\footnote{%
This paper collects into one place my replies to the questions posed by
Maximilian Schlosshauer in his interview volume about the foundations of
quantum mechanics, \textit{Elegance and Enigma: The Quantum Interviews}, ed.
M. Schlosshauer (Springer, 2011).}}

\bigskip

\bigskip

\bigskip

Antony Valentini

\textit{Department of Physics and Astronomy,}

\textit{Clemson University, Kinard Laboratory,}

\textit{Clemson, SC 29634-0978, USA.}

\bigskip

\bigskip

\bigskip
\end{center}

\noindent \textbf{Q1. What first stimulated your interest in the foundations
of quantum mechanics?}

It so happens that I studied general relativity before I studied quantum
theory. This seems to have affected my reaction to the latter. It surprised
me that quantum theory uses the same elementary degrees of freedom as
classical physics does -- position, momentum, angular momentum, and so on.
Nor are any radically new degrees of freedom introduced (with the possible
exception of spin). Classical variables are promoted to operators. The
values are restricted to eigenvalues. But the set of elementary variables is
the same as in classical physics. General relativity, in contrast, abandons
Newtonian gravitational force, and introduces the new concept of curved
spacetime. The basic ontology changes. I had expected quantum theory to be
in a sense much more novel than it is. But the basic variables are
unchanged, only subject to an operator calculus that is itself constructed
by analogy with classical physics. In this sense, the theory talks about
nonclassical systems as if they were still classical.

My experience with general relativity had led me to expect that quantum
theory would be based on, as I would now put it, a different ontology from
that of classical physics. I was disappointed that there was no new,
distinctively quantum entity. There is the wave function, of course, but it
seemed just to give probabilities for values to be taken by the usual set of
classical variables. I was also impressed by the conceptual leap that
Faraday had made in developing the concept of the electromagnetic field.
This was a novel and non-Newtonian ontology, a fundamentally new kind of
thing. What comparable new concept did quantum theory introduce? None that I
could see.

Then I remember the first time I encountered an explicit statement of the
collapse postulate. The book I read said something about how when a system
is observed it is `thrown into' an eigenstate. How could merely looking at
something make it jump around like that?

As an undergraduate, I read Ballentine's famous 1970 review of the
statistical interpretation, and was impressed to see so many muddles cleared
up. However, it pointed to something deeper, along the lines of hidden
variables, and by itself could not be satisfactory. I also read a bit about
de Broglie-Bohm theory, but unfortunately the papers I read contained
obvious mistakes and this led me to think the theory was wrong or at best
incomplete.

Later on, during some years I spent outside academia after graduating, I
became deeply impressed by quantum nonlocality, and in particular by the
puzzle of why we can't use it for instantaneous signalling. In the 1980s,
some people were still proposing ways to use EPR correlations for
signalling, and of course it would always be shown by someone else that the
proposals were wrong. The no-cloning theorem, for example, arose in response
to such a proposal. I couldn't shake off the feeling that there was
something nonlocal going on behind the scenes, and that quantum uncertainty
noise was preventing us from seeing it. By means of Bell's theorem, we could
deduce that the nonlocality was there, but the uncertainty principle stopped
us from using it for signalling. This was my strong impression. Shimony
referred to this sort of thing as a `peaceful co-existence' between
relativity and quantum theory. To me, it seemed like a dark and uneasy
conspiracy that cried out for an explanation.

I started to form the vague idea that the hidden-variable level must be in a
state of some sort of statistical equilibrium, in which the nonlocal effects
average to zero. In ordinary statistical mechanics, thermal equilibrium
yields finely-tuned relations such as detailed balancing, the
fluctuation-dissipation theorem, and so on. I was also fascinated by an
analogy with Maxwell's demon, who is unable to sort fast and slow gas
molecules only because he is in thermal equilibrium with the gas and is
therefore subject to the same thermal noise as the molecules themselves. I
started to think that something similar must be going on at the
hidden-variable level -- that we are unable to control the details of hidden
variables because we are ourselves stuck in an equilibrium state. I had
vague ideas about a hypothetical `subquantum demon', who could predict
outcomes of spin measurements more accurately than quantum theory allows,
and who could thereby use EPR correlations for nonlocal signalling. But I
didn't have a real theory. This must have been around 1988. It was only when
I studied de Broglie-Bohm theory properly, in early 1990, that I saw a
concrete way to realise the idea.

\medskip \noindent \textbf{Q2. What are the most pressing problems in the
foundations of quantum mechanics today?} \\[-0.2cm]

The interpretation of quantum mechanics is a wide open question, so we can't
say in advance what the most pressing problems are. As the history of
physics shows, it's only in hindsight that one can say who was looking in
the right direction. What's important is that we leave the smokescreen of
the Copenhagen interpretation well behind us, and that talented and
knowledgeable people think hard about this subject from a realist
perspective.

Instead of answering the question, I can offer a list of things I'd like to
see done in the near future, as they seem important as far as I can tell.

It would be good if the ongoing controversy over the consistency of the
Everett intepretation could be settled. It would be helpful to know if that
theory really makes sense (on its own terms) or not. It would also be good
to see further experiments searching for wave function collapse. More
generally, I'd like to see more experiments that test quantum theory in
genuinely new domains -- as in the recent three-slit experiment.

In modern theoretical physics, there are a number of important issues that
deserve more attention from a foundations perspective, such as the question
of Hawking information loss in black holes, and the problem of time in
quantum gravity. The description of the quantum-to-classical transition in
the early universe also deserves more foundational scrutiny.

As for my own current line of research -- which focusses on the possibility
of nonequilibrium violations of quantum theory, in de Broglie-Bohm theory
and in deterministic hidden-variables theories generally -- there are some
outstanding issues that need a lot more work. One is the need for more
detailed calculations and numerical simulations of relaxation to quantum
equilibrium in the early universe, with the aim of obtaining precise
predictions of where residual nonequilibrium violations of quantum theory
might be found today, for example in the cosmic microwave background or in
relic cosmological particles. My work so far points in the direction of
super-Hubble wavelengths as the area to look at, but much more remains to be
done. I have also made some proposals to the effect that Hawking radiation
could consist of nonequilibrium particles that violate the Born rule, in a
way that might avoid information loss, and there are a host of theoretical
questions to be investigated to develop that proposal further.

Finally, there is the important general question of whether it's possible to
construct a reasonable hidden-variables theory without an ontological wave
function. De Broglie-Bohm theory has several features that have been shown
to be common to all hidden-variables theories (under some reasonable
assumptions): nonlocality, contextuality, and nonequilibrium superluminal
signalling. De Broglie-Bohm theory also has the feature of an ontological
wave function, and it would be good to know if this is another common
feature of hidden-variables theories or not. Alberto Montina has worked on
this recently, but more needs to be done.

\medskip \noindent \textbf{Q3. What interpretive program can make the best
sense of quantum mechanics, and why?} \\[-0.2cm]

In my view, the de Broglie-Bohm interpretation -- or pilot-wave theory, as
de Broglie originally called it -- presents several deeply attractive
features, in addition to the obvious merits of being realistic and
deterministic.

First and foremost, as I said in my answer to Question 1, before I studied
de Broglie-Bohm theory properly, I was very puzzled by why we can't use
quantum nonlocality for signalling. It was as if there was some sort of
conspiracy at work in the laws of physics. To explain this, I had drawn the
conclusion that we were stuck in some sort of statistical equilibrium state
whereby uncertainty noise happens to mask the underlying nonlocal effects (a
kind of equilibrium `balancing' condition, as I had vaguely thought of it).
When I studied pilot-wave theory and saw that it was a consistent theory, I
was amazed to see that it provided a precise realisation of what I had been
looking for: it was a nonlocal theory, for which the nonlocality was washed
out or averaged to zero in the state of `quantum equilibrium' -- the state
in which the hidden configurations have a Born-rule distribution. Other
people working on the theory usually took the Born probability rule as an
axiom, alongside the equations of motion, but to me it seemed obvious that
the theory should be considered for arbitrary distributions. I was able to
show that such distributions give rise to nonlocal signals at the
statistical level. I also proved an analogue of the classical
coarse-graining \textit{H}-theorem, which gave a general understanding of
how evolution towards equilibrium occurs for an isolated system, as has
since been confirmed by numerical simulations. It seemed natural to me to
suppose that relaxation to equilibrium had taken place in the remote past,
presumably soon after the big bang. The nonlocality associated with early
nonequilibrium might then explain why the early universe was so homogeneous.
More importantly, the puzzle of why we can't use quantum nonlocality for
signalling could be given a simple answer: it's a peculiarity of the fact
that we happen to be stuck in an equilibrium state. There's no conspiracy in
the laws of physics, we are simply trapped in a special state with special
properties. Both signal-locality and the uncertainty principle could be
shown to be contingencies of equilibrium.\footnote{%
A. Valentini, `Signal-locality, uncertainty, and the subquantum \textit{H}%
-Theorem. I and II', \textit{Phys. Lett. A} \textbf{156}, 5; \textbf{158}, 1
(1991).}

This viewpoint opens up the possibility of a new and wider `nonequilibrium'
physics, in which superluminal signalling is possible and the uncertainty
principle can be circumvented. Relativity, too, is violated in this new
physics, which contains a notion of absolute simultaneity associated with a
preferred state of rest. Nonequilibrium particles that violate the Born rule
could exist today, perhaps in the form of relic particles from the very
early universe -- particles that decoupled before they had time to relax
completely to equilibrium. I've also speculated that nonequilibrium
particles might be generated by evaporating black holes, on the grounds that
their states could then carry more information and possibly avoid the
Hawking information loss.

I find pilot-wave theory attractive in another respect. In de Broglie's
original formulation, it is a radically non-Newtonian theory, with the
extraordinary feature of a wave in configuration space, that determines the 
\textit{velocities} of systems instead of their accelerations. It's an
Aristotelian dynamics, with a natural Aristotelian kinematics. Nonlocality
in ordinary space is explained as an effect of the configuration-space
dynamics. This seems to me much more like the radical conceptual shift I had
expected when I first studied quantum theory (see my answer to Question 1).
We throw out classical forces and classical dynamics, introduce a new and
radical entity (the pilot wave in configuration space), and construct a new
theory of motion. That's the sort of thing that Faraday and Maxwell did with
the electromagnetic field in the nineteenth century, and that Einstein did
with gravity in 1915. And that's what de Broglie did in the 1920s.

In 1923, de Broglie concluded that, when a particle is diffracted by an
obstacle without touching it, the non-rectilinear motion violates Newton's
first law. De Broglie proposed a new form of dynamics based on velocities,
in which particle motions were guided by waves, in a way that unified the
variational principles of Maupertuis and Fermat. But de Broglie's
achievement went unnoticed. He is remembered chiefly for the relation
between momentum and wavelength, but that was merely a by-product of his new
dynamics. Even those who work on his theory often fail to credit de Broglie
-- who in 1927 had the full many-body dynamics in configuration space, not
just the one-body theory as is often claimed.\footnote{%
G. Bacciagaluppi and A. Valentini, \textit{Quantum Theory at the Crossroads:
Reconsidering the 1927 Solvay Conference} (Cambridge University Press, 2009).%
}

Unfortunately, in 1952, Bohm presented the theory in a pseudo-Newtonian
form, based on acceleration and the quantum potential, which made it look
much more like classical physics than it really was. Bohm's important
contribution was to show how the theory accounts for the general quantum
theory of measurement. But I never took Bohm's version of the dynamics
seriously. It looks artificial, like writing classical general relativity in
terms of flat spacetime with a tensor field that distorts rods and clocks.
It can be done, but it's not a natural language to use. De Broglie's
original dynamics, which Bell used and advertised, seems much more
appropriate. In recent work with Samuel Colin, we have shown that Bohm's
dynamics is actually unstable, in the sense that non-standard momentum
distributions (which are allowed in Bohm's dynamics but not in de Broglie's)
do not relax to quantum equilibrium. So my preference for de Broglie's
dynamics is no longer merely a question of taste. I think Bohm's dynamics is
actually untenable.

De Broglie's remarkable work in the 1920s remains mostly unknown, even among
historians. In my view, in certain key respects, he understood the
fundamental dynamics better than Bohm did. Matters are further confused by
some who refer to de Broglie's dynamics as `Bohmian mechanics'. For a proper
understanding of the theory, it helps to know how and why de Broglie
constructed it in the 1920s -- instead of thinking of it anachronistically
in terms of a `completion' of modern quantum theory. In my view, the theory
is still widely misunderstood, even by some of its most fervent supporters,
partly because de Broglie's original work is still being ignored.
Associating the theory primarily with Bohm is not only wrong as regards
credit, it also deprives us of de Broglie's insights.

While de Broglie's dynamics is attractive in itself, for me it has always
been first and foremost a means to provide a concrete model of the idea I
had about physicists being trapped in an equilibrium state that hides
nonlocality. I thought from the outset that the essence of this idea would
hold in \textit{any} reasonable deterministic hidden-variables theory -- as
I eventually showed explicitly. Now, at present, we don't know what the true
hidden-variables theory is. Pilot-wave theory might be right or
approximately right; of course, it could also be quite wrong. I think it's a
worthy guess, as it contains a number of features, such as nonlocality, that
are known to be generally true for hidden-variables theories. But to find
out what the correct theory is, we'll need an empirical window.

I tend to compare pilot-wave theory with the early models of the kinetic
theory of gases, in which molecules were hard spheres. That was the simplest
assumption to make, and it was a good strategy to develop the resulting
theory as far as possible, until things like the explanation of Brownian
motion gave us an empirical window onto the world of atoms. Similarly, I
hope that developing pilot-wave theory to its logical conclusions will lead
to an empirical window onto the world of hidden variables.

I wouldn't be willing to bet a huge sum that the details of pilot-wave
dynamics are correct. However, I \textit{would} be willing to bet a
considerable sum that there is a nonlocal hidden-variables theory behind
quantum mechanics, and that the only reason we can't send superluminal
signals today is because we're trapped in a state in which the hidden
variables have an equilibrium distribution. Locality and the uncertainty
principle are not laws, they are merely peculiarities of equilibrium.
Quantum theory is a special case of a much wider, nonequilibrium physics, in
which nonlocal signalling is possible and the uncertainty principle can be
beaten. I think this is likely to be true. And it's a good scientific rule
of thumb to say that, if the laws of physics permit something to happen,
then it will happen somewhere. So, I expect that nonequilibrium violations
of quantum theory will eventually be found. When they are found, and we are
able to see our way through the fog of quantum noise, what will we find?
Will we see trajectories obeying pilot-wave dynamics? Maybe, maybe not. But
we will, I think, see a nonlocal world radically different from the world
we're familiar with. And we will realise how misled we've been all this
time, wrongly thinking that the Born rule and its associated features are
fundamental when they are not.

\medskip \noindent \textbf{Q4. What are quantum states?}

In my view, de Broglie's pilot wave is a new kind of causal agent, a
radically new kind of physical entity grounded in configuration space. To
understand it, it's helpful to examine the historical parallel with two
other physical entities that seemed mysterious when first introduced:
Newton's concept of gravitational attraction-at-a-distance, and Faraday's
concept of field.

Before the acceptance of Newtonian gravity, it seemed to many that
scientific explanation should be reduced to Cartesian action-by-contact. If
one body appeared to act on another at a distance, there must be an
intervening medium that transmits the force through local action by contact.
In Newtonian gravity, instead, a massive body can act directly on another at
a distance, through empty space. Even Newton himself had difficulty with the
idea, and continued to seek a deeper explanation for gravity in terms of an
aetherial medium filling space. The concept of gravitational attraction
arose by a process of abstraction, in which the conceptual scaffolding of a
Cartesian medium was thrown away. A similar step occurred in the nineteenth
century, when Faraday introduced the concept of field. Faraday looked at the
pattern of iron filings around a bar magnet, and started to think that the
pattern would exist even if the filings were taken away. It's hard for us
today to appreciate what a conceptual leap that was. Remember, at that time,
a force was understood to be present when a mass accelerates. In empty
space, where there are no masses, how could a force exist all by itself? But
Faraday believed that the magnetic `lines of force' seen in patterns of iron
filings existed in their own right, even when the filings were absent. We
eventually got used to the idea of forces existing and even propagating in
empty space, where there are no masses or charges. Again, the concept arose
by a process of abstraction. By abtracting away the iron filings, we're left
with the concept of fields in their own right.

Now, in my view, history repeated itself in 1927, when de Broglie introduced
the concept of a pilot wave in configuration space. Again, it arose by a
process of abstraction. In particular, early in 1927 he was struggling with
a model he had of a system of particles as singularities of coupled fields
in three-space. He was trying to show that the singularities would follow
the guidance equation, which was supposed to emerge as an effective
description of the motions, which were ultimately generated by the
complicated coupled field equations. But he saw that, as a provisional
theory, he could simply take the guidance equation with the pilot wave in
configuration space, and forget about the underlying model -- just as Newton
did with gravity, and just as Faraday and others did with electromagnetism.
The difference, though, is that while we all recognise Newton and Faraday
for their achievements, most physicists and historians simply don't know
what de Broglie really did in the 1920s. I believe that, in 1927, de Broglie
introduced a fundamentally new entity into physics, but to this day the
world hasn't really noticed, and even those who are interested in his theory
do not properly understand what de Broglie did. And of course, to complete
the analogy, particularly with Newton, de Broglie himself was uncomfortable
with the idea, and thought it should emerge as an effective theory along the
lines he had been considering -- a view he returned to in later life. He
never really believed the pilot wave in configuration space was fundamental,
just as Newton never believed that his theory of gravity was fundamental.
Still, Newton's concept lasted for more than two hundred years.

In my view, de Broglie's pilot-wave concept deserves to be taken more
seriously. It might turn out to be useful-but-wrong, like Newton's concept
of gravitational attraction-at-a-distance. Or, it might turn out to be an
essential new concept -- as happened with Faraday's concept of field, which
survives even in quantum field theory. In any case, at present, I would
suggest that the wave function is a new kind of causal agent, as new and
radical as was Faraday's concept of field, but which for historical reasons
has not been recognised.

To see how radical it is, consider the contrast with the idea of a field in
space. An ordinary field can be probed using a test particle. An electric
field, for example, can be measured by introducing an infinitesimal test
charge and watching it accelerate. But if we try to do this for the pilot
wave, we find that introducing a test particle actually increases the
dimension of the configuration space on which the wave is defined. There is
no such thing as a test particle for the pilot wave. So it's not comparable
to ordinary fields, something that Bohm didn't really appreciate in 1952,
but which de Broglie understood in 1927.

We may need to learn to think in terms of pilot waves, just as earlier
generations learned to think in terms of fields, without a conceptual
scaffolding based on more primitive notions. In particular, the pilot wave
in configuration space provides a natural understanding of nonlocality in
3-space.

Finally, I should comment on the notion of pilot wave for the whole
universe. Aside from gaps in our understanding of quantum gravity, I see
nothing problematic there. Some people claim that, because there is `only
one universe', the universal pilot wave cannot be contingent, and must
instead be law-like. But that argument is spurious. In our current
understanding of cosmology, the intergalactic magnetic field is not
determined by physical laws -- it is contingent, in the sense that, for all
we know, the configuration of that field could have been different. The same
goes for the spacetime geometry of the universe as a whole. And the same can
be said of the universal pilot wave.

\medskip \noindent \textbf{Q5. Does quantum mechanics imply irreducible
randomness in nature?} \\[-0.2cm]

Certainly not. There is at least one formulation of quantum mechanics -- the
pilot-wave theory of de Broglie and Bohm -- that has no such randomness,
therefore the conclusion cannot be drawn.

Pilot-wave theory has been extended to cover high-energy physics, with
different approaches taken by different authors. In the best approach, in my
view, bosons are described in terms of c-number fields, while fermions are
particles with a pilot wave obeying the many-body Dirac equation. For
fermions, we have to take the Dirac sea seriously. Even the vacuum is full
of particles. This model was proposed by Bohm and Hiley, and its relation to
quantum field theory has been clarified by Colin and Struyve. This, together
with the bosonic field theory, provides a completely deterministic theory of
high-energy physics, including processes such as pair creation. Some workers
have proposed models of fermions in which pair creation has a fundamentally
stochastic element, but those models are unnecessarily cumbersome. It would
be odd if pair creation forced indeterminism upon us. It is in fact
straightforward to construct a completely deterministic theory of such
processes.

Within pilot-wave theory, it has been claimed that the Born rule has a
fundamental status as a preferred measure of `typicality' for the initial
configuration of the universe. If this were so, in practice we would always
be stuck with randomness for subsystems. But that argument inserts the Born
rule by hand at the initial time. As I've said (in my reply to Question 3),
the theory certainly allows for `nonequilibrium' violations of the Born
rule. Such violations for subsystems are `untypical' with respect to the
global Born-rule measure. But to claim that they are therefore intrinsically
unlikely is circular, because such violations are readily shown to be
`typical' with respect to non-Born-rule measures. Also, I don't see a
difference between `typicality' and `probability'. To say that we will
always have Born-rule randomness in practice, as some have argued, is in my
view mistaken. There is no good reason to believe that. On the contrary, if
one takes the theory seriously, it suggests that nonequilibrium will
eventually be found somewhere, as I urged in my reply to Question 3.

I have shown that quantum nonequilibrium systems could be used to perform
`subquantum measurements' on ordinary systems.\footnote{%
A. Valentini, `Subquantum information and computation', \textit{Pramana --
J. Phys}. \textbf{59}, 269 (2002).} These are measurements that violate the
uncertainty principle and other standard quantum constraints. An extreme
nonequilibrium ensemble, with arbitrarily small dispersion, could be used to
perform analogues of the ideal, non-disturbing measurements familiar from
classical physics. These would allow us to track the trajectories without
disturbing the wave function, and to predict the future in ways that are not
allowed by quantum theory. In other words, quantum randomness could be
circumvented in the laboratory, if we possessed such nonequilibrium systems.
And it's conceivable that such systems could exist today, in the form of
relic particles from the very early universe (see my answer to Question 11).

\medskip \noindent \textbf{Q6. Quantum probabilities: subjective or
objective?}

From a de Broglie-Bohm point of view, the situation is more or less the same
as in classical statistical mechanics. At the fundamental level, there is no
such thing as probability. The universe contains a huge number of degrees of
freedom evolving according to deterministic equations of motion. In
principle, that's all there is to it. In practice, for large numbers of
similar and approximately independent systems, it's useful to work with a
distribution of configurations, and to consider the theoretical limit of an
infinite ensemble. This is only a practical tool. The interpretation of that
distribution depends on what approach you take to probability theory. This
leads to interesting questions in the foundations of probability theory, but
those questions have nothing particularly to do with pilot-wave theory, they
arise in a similar way in ordinary classical statistical mechanics.

Because the theory is fundamentally deterministic, it may seem natural to
characterise a probabilistic description as `subjective'. On the other hand,
the statistics we see in the lab are properties of the actual configuration
of our universe, so in that sense they are `objective'.

Questions about the foundations of probability theory arise not only in
statistical mechanics, but in any application of probability theory or
statistical inference, for example to genetic populations on earth or to the
distribution of galaxies in deep space. De Broglie-Bohm theory has nothing
new to add to such debates, and so I try to avoid them.

We should avoid getting distracted by such questions in a de Broglie-Bohm
context, when the focus should be on finding evidence for the details of the
underlying dynamics. There's a parallel with atomic physics in the late
nineteenth century. Boltzmann's central belief was that everything was made
of atoms, and that macroscopic physics could be reduced to atomic physics.
In retrospect, it's a pity that he got distracted by controversies relating
to the foundations of probability theory, as well as by questions concerning
time reversal, and so on, when the priority was to find evidence for atoms.
Similarly, while I agree that conceptual questions about the meaning of
probability are interesting, I think that in the context of pilot-wave
theory they are at best distracting us from more important issues, and at
worst obscuring the physics of the theory -- which is fundamentally a
nonequilibrium physics that violates quantum mechanics.

As an example of the sort of thing I mean, some people in quantum
foundations talk as if it is problematic to consider probabilities for the
`whole universe'. And yet, cosmologists not only do so every day, they are
also busy testing primordial probabilities experimentally by measuring
temperature anisotropies in the cosmic microwave background. By making
statistical assumptions about a theoretical `ensemble of universes',
cosmologists are able to test probabilities in the early universe, such as
those predicted by quantum field theory for vacuum fluctuations during
inflation. One can question what the ensemble of universes refers to. Is it
a subjective probability distribution? Or, is the universe we see in fact a
member of a huge and perhaps infinite ensemble, as is the case in theories
of eternal inflation? Those are interesting questions, but only tangentially
related to the ongoing experimental tests. This point is related to my
critique, in my answer to Question 4, of supposed problems with contingency
for the universal wave function. I don't see why people working in quantum
foundations should worry about such matters in a cosmological context, when
cosmologists do not.

\medskip \noindent \textbf{Q7. The quantum measurement problem: serious
roadblock or dissolvable pseudo-issue?} \\[-0.2cm]

The measurement problem is often stated as the problem of the interpretation
of a quantum superposition, such as Schr\"{o}dinger's cat. That is
inaccurate and misleading. Among other things, it allows for the facile
response that the wave function refers only to a statistical ensemble.
However, while such a `statistical' or `epistemic' interpretation might turn
out to be correct, it does \textit{not} solve the true and deep measurement
problem, which is the problem of what happens to macroscopic realism at
microscopic scales. In quantum physics, we have definite states of reality
at the macroscopic level but not at the microscopic level. There is no
precisely-defined boundary between these two domains. Therefore, standard
quantum theory is fundamentally ill-defined.

An apparatus pointer in the lab, for example, points in a definite
direction. Particles, on the other hand, generally have indefinite
positions. How many particles are required to make a `macroscopic' pointer?
How many are required, to cross the line from microscopic fuzziness to
macroscopic definiteness? There is no precise dividing line between the
microscopic and the macroscopic. And all macroscopic equipment is built out
of microscopic systems. How do definite states arise from indefinite ones?
There's a temptation here to think in terms of emergence, but there can be
no continuous transition from indefinite to definite states of reality.
Either something exists or it does not.

Some people say that there are `degrees of reality', that one object can be
`more real' or `less real' than another. One sometimes hears physicists ask
if a rock somewhere out in deep space is real when no one is looking at it.
But such talk misunderstands the nature of the word `real'. When we say that
`X is real', we simply mean that `X exists'. If I say there is a rock out in
deep space with no one looking at it, I have already stated that the rock
exists, that is, I've already said the rock is real. To then suggest that
perhaps the rock is not real, because no one is looking at it, is to
contradict oneself. An analytical philosopher would probably convey this
point by saying that `real' is a quantifier, not a property. But the point
is simple enough, and is the basis of much elementary reasoning, both in
physics and outside of it.

Others try to evade the measurement problem by claiming that the usual
notion of ontology depends on a `God's-eye view' of the world. But that is
mistaken. For example, if a piece of macroscopic apparatus, with a dial and
a pointer, has a particular setting and pointer reading, this is not
dependent on anyone's (or God's) `viewpoint', it is simply a fact about the
dial and pointer. Facts require neither a human observer nor a deity. They
are facts, whether or not we or God or whoever is there to know them.

Other attempts to avoid the issue include recourse to nonclassical logic.
One can invent a mathematical structure that violates some rules of formal
logic and \textit{call} it a `logic'. However, everyone still uses so-called
classical logic in order to reason and argue. I think there's a misuse of
words here. Logic is logic. If there's a contradiction, for example in a
thought experiment such as that of Wigner's friend, then it won't do to
dismiss it as a failure of classical logic. The contradiction comes from
clear thinking, and requires a clear answer.

Finally, some people say that the concept of objective reality must be
abandoned even at the macroscopic level. But we each know that we exist, and
if we have any sense we will know that other minds exist as well. There is a
world out there, containing other human beings, as well as things like
tables and chairs, and pieces of equipment with dials and pointers.

To make quantum mechanics a precise theory, we must posit the existence of
something that extends into the microscopic domain. This can and has been
done in various ways, involving hidden variables, or many worlds, or
collapse theories. It remains to be seen which, if any, of these proposals
are correct.

To suggest that the measurement problem is a pseudo-issue is to say that the
simple question `what is real?', or equivalently `what exists?', does not
require an answer. When people say that, they are being inconsistent,
because they themselves talk about `what is real' or `what exists' when it
comes to things like the outcomes of experiments in their laboratories, or
what car they own. Everyone uses the notion of definite states of objective
reality at the macroscopic level, including in the laboratory -- when it is
asserted, for example, that we really did find a certain wavy pattern of
dots on a photographic film in an interference experiment. It's only at the
microscopic or quantum level that there is controversy over what is real. To
say that we don't need a notion of microscopic reality at all, while at the
same time using a notion of macroscopic reality whenever one describes an
experiment, is to ignore the self-evident ambiguity in the dividing line
between microscopic and macroscopic, and to ignore the resulting
self-evident ambiguity in what one is saying.

\medskip \noindent \textbf{Q8. What do the experimentally observed
violations of Bell's inequalities tell us about nature?} \\[-0.2cm]

The observed violations of Bell's inequality tell us that locality is
violated -- if we assume that there is no backwards causation and that there
are not many worlds.

There is a widespread misunderstanding that Bell's theorem assumes
determinism or the existence of hidden variables. In fact, Bell's original
1964 argument had two parts. The first part uses the EPR argument to show
that, if locality is assumed, then quantum outcomes must be determined in
advance. The second part takes this deduction as a starting point, and goes
on to prove the famous inequality. As Bell himself emphasised, determinism
is not assumed, it is deduced from locality in the first part of the
argument. So the contradiction is not only between hidden variables and
locality. There is a contradiction between quantum theory itself and
locality (in the absence of backwards causation and many worlds).

If we allow for backwards causation, so that future apparatus settings can
affect systems in the past, then it seems that nonlocality is not required.
It's unfortunate that very little work has been done developing such models
to cover a broad range of physics, so we don't know if plausible and
attractive theories along these lines exist. As for many worlds, it's a
possibility -- in my view unlikely, but possible -- though whether that
theory is well-defined remains controversial.

\medskip \noindent \textbf{Q9. What contributions to the foundations of
quantum mechanics have or may come from quantum information theory? What
notion of `information' could serve as a rigorous basis for progress in
foundations?} \\[-0.2cm]

In my view, with the rise of quantum information theory in the 1990s, the
subject of quantum foundations was set back by at least 20 years. There are,
however, some issues that need to be clearly distinguished.

First of all, quantum information theory is just quantum mechanics applied
to certain practical problems. Nothing new is said about ontology, and the
usual ambiguities remain. No attempt is made even to address the measurement
problem.

On the other hand, quantum information theory has emphasised some aspects of
quantum theory that had been unduly neglected. In particular, entanglement,
peculiarities of the tensor product structure of Hilbert space, and general
properties of unitary evolution such as the no-cloning theorem. What these
features have in common is that they don't depend on details of the system
or on what its Hamiltonian happens to be. Some people find this exciting.
But in fact, systems do consist of particles and fields, and these propagate
in spacetime, and there are various symmetries associated with conservation
laws, and so on, and all this remains the basic stuff of physics. It should
also be remembered that entanglement as the fundamental new feature of
quantum physics was discussed by Schr\"{o}dinger as long ago as 1935,
quantum cryptography was anticipated by Wiesner around 1969, and explicit
statements of the no-cloning theorem date from 1982. Much of the basic and
truly important material is not as novel as is often claimed. What has
really happened is that these features have turned out to be of
technological interest, and the resulting outpouring of funding has
generated a huge bandwagon.

As far as fundamental physics is concerned, I see a useful parallel with
what happened in general relativity in the 1960s, when people discovered
that some important deductions could be made purely on the basis of
geometrical arguments, without invoking the details of Einstein's field
equations. I mean results like the singularity theorems of Penrose and
Hawking. Modern textbooks on general relativity include a chapter on such
geometrical methods -- containing, in particular, a few key results such as
the singularity theorems, and a few useful theorems about global
hyperbolicity and causal structure. But still, most of what we know and
understand about Einsteinian gravity comes from analysis of the field
equations. Now, the parallel with quantum information theory is clear. It
was realised that some important deductions could be made purely from
geometrical or kinematical properties of unitary evolution in Hilbert space.
The details of the Schr\"{o}dinger equation or Hamiltonian didn't matter. It
will soon be standard for introductory textbooks on quantum mechanics to
contain a chapter giving a few key results such as the no-cloning theorem
and one or two useful theorems about entanglement. But still, most of what
we know about quantum physics comes from analysis of the theory applied to
concrete systems of electrons, photons, atoms, and so on, and the detailed
structure of the Hamiltonian is of central importance.

But the real damage that has been done is in reviving the misguided idea
that physics is only about macroscopic operations and observations. A sort
of `neo-Copenhagen' attitude has arisen, with the word `information' playing
a role similar to the older word `observation'. The usual ambiguities
remain. Macroscopic equipment with its definite ontological states plays a
fundamental role, while no ontology is provided at the microscopic level,
and with no heed paid to the lack of a clear dividing line between those two
levels. The measurement problem is simply not addressed.

It is sometimes claimed that `information' is a fundamental new concept. But
`information' is synonomous with `knowledge about something'. What is the
knowledge about? If it is only about macroscopic instrument readings, then
it is not knowledge of anything fundamental.

I see quantum information theory as also analogous to thermodynamics. In the
late nineteenth century, some people thought that they had found a new
approach to physics, that focussed on the production, transmission and use
of energy, based on general principles that didn't depend on details of the
system. In retrospect, of course, gases and liquids are made out of atoms
and molecules, and their macroscopic behaviour is not fundamental but
emergent. Nowadays, some people claim that physics is about the production,
transmission and use of information, based on general principles that don't
depend on details of the system. But again, the systems we see are built out
of microscopic entities, and the behaviour of macroscopic instruments is not
fundamental but emergent.

\medskip \noindent \textbf{Q10. How can the foundations of quantum mechanics
benefit from approaches that reconstruct quantum mechanics from fundamental
principles? Can reconstruction reduce the need for interpretation?} \\[-0.2cm%
]

I don't think quantum mechanics is a fundamental theory. It's ambiguous. And
it's ambiguous because it lacks a microscopic ontology. Any reconstruction
that does not provide such an ontology will remain ambiguous and therefore
not fundamental. We see this in work over the past decade or so,
reconstructing quantum theory from various operational axioms. Those axioms
refer only to outcomes of experiments performed with macroscopic equipment.
They provide constraints on the statistical properties of those outcomes.
This may be of some interest, but only up to a point. Nothing is said about
fundamental ontology. Pieces of macroscopic equipment are treated as if they
were fundamental or elementary objects, when in reality they are emergent
objects built out of atoms, particles, and fields. The pieces of equipment
are assigned definite ontological states -- the pointers point in definite
directions, the knobs and dials on the apparatus have definite readings, and
so on -- while microscopic systems are not. Nothing is said about the
dividing line between the definite macroscopic world and the indefinite
microscopic world. Therefore, these operational approaches remain
fundamentally vague. They do not attempt to address the measurement problem,
therefore they are of limited interest.

Pilot-wave dynamics, in contrast, does provide a reconstruction of quantum
mechanics in terms of a fundamental ontology that is equally valid at the
macroscopic and microscopic levels. There are two simple equations of
motion, de Broglie's guidance equation and Schr\"{o}dinger's wave equation.
As with other fundamental equations of physics -- such as Maxwell's
equations or Einstein's field equations -- one can try to motivate these
equations on the basis of simple physical principles. In the early 1920s, de
Broglie motivated his guidance equation as a way to unify the principles of
Maupertuis and Fermat. The Schr\"{o}dinger equation is the simplest wave
equation that respects the nonrelativistic dispersion relations. Thus,
simple physical principles suggest two general equations of motion, which --
if an initial Born-rule distribution is assumed -- provide a complete and
unambiguous reconstruction of quantum mechanics as an emergent equilibrium
phenomenology. Though I wouldn't put too much emphasis on the motivating
principles. At the end of the day, the basis of the theory is the equations
themselves.

As for reducing the need for interpretation, that happens only if we provide
an ontology. The question being asked probably refers to reconstruction
along operationalist lines, which has become fashionable in recent years. As
I've explained, that work does not even attempt to address the measurement
problem. People often draw an analogy with special relativity. In 1905,
Einstein gave an operational treatment based on macroscopic rods, clocks,
and light beams, and he derived the Lorentz transformation from a small
number of simple principles. Current work in operational quantum theory
seeks to emulate that. In my view, Einstein's famous 1905 paper is the
historical source of a serious mistake, whereby macroscopic equipment is
given a fundamental role -- a mistake that was repeated by Bohr, Heisenberg
and others in the 1920s, with catastrophic consequences. Like any other
piece of macroscopic equipment, rods and clocks are not elementary systems,
they are emergent objects built out of particles and fields. Our modern
understanding of Lorentz invariance, commonly described in textbooks on
high-energy physics and quantum field theory, boils down to having a
Lagrangian density that is a Lorentz scalar. It's a symmetry of the basic
equations. There is no mention of rods and clocks, or of any principle about
the speed of light -- the photon could, after all, turn out to have a small
mass and move at slightly sub-luminal speeds. I think Einstein's 1905 paper
was deeply damaging, and continues to be so. Nor was it necessary. The
structure of special relativity was independently derived by Poincar\'{e} in
1905, by generalising the Lorentz invariance of Maxwell's equations to all
the laws of nature -- precisely the approach that a modern particle
physicist would have taken. I see little to emulate in Einstein's first
paper on special relativity, and much to deplore. Einstein himself deeply
regretted the operational fashion he started in that paper.

I see different formulations of operational quantum theory as analogous to
different formulations of thermodynamics. People can argue over whether
Kelvin's formulation of the second law is better than that of Clausius, or
whether Carath\'{e}odory's geometrical approach is to be preferred. But in
the end, they are merely talking about different axiomatisations of the same
phenomenological theory, none of which bears on the burning issue of
fundamental ontology.

\medskip \noindent \textbf{Q11. If you could choose one experiment,
regardless of its current technical feasibility, to help answer a
foundational question, which one would it be?}

There are at least five different experiments that I would be keen to do,
all of them involving tests of the Born probability rule.\footnote{%
A. Valentini, `Astrophysical and cosmological tests of quantum theory', 
\textit{J. Phys. A: Math. Theor}. \textbf{40}, 3285 (2007).}

First, I would like to test the Born rule for particles that have been
emitted by an evaporating black hole. This could be done at least in
principle, should we discover primordial black holes, left over from the
early universe, that are currently evaporating. I would also like to test
the Born rule for particles that are entangled with partners that have
fallen behind the event horizon of a black hole. This might be possible, if
we find appropriate atomic cascade emissions taking place naturally in the
neighbourhood of a supermassive black hole. I would also be keen on testing
the Born rule for any kind of particle at the Planck scale. The motivation
for these three experiments is my suggestion that the quantum equilibrium
state might become unstable in the presence of gravity.

Another worthwhile place to look, in my view, is in the neighbourhood of
nodes of the wave function, where the de Broglie-Bohm dynamics breaks down.

The fifth experiment I'm keen on is to test the Born rule for relic
cosmological particles that decoupled (at early times) when their
wavelengths were larger than the instantaneous Hubble radius. By analysis of
the relaxation process on expanding space, I've shown that relaxation can be
suppressed at super-Hubble wavelengths, so it's possible that such particles
never underwent relaxation -- they could still exist in our universe today,
and violate the usual rules of quantum mechanics. Specifically, I would
suggest testing their polarisation probabilities, for the particles
themselves or for their decay products, and to search for violations of
Malus' law. Since I have to choose only one experiment, let it be this last
one.

\medskip \noindent \textbf{Q12. If you have a preferred interpretation of
quantum mechanics, what would it take to make you switch sides?}

The observation of spontaneous collapse would of course make me switch to
collapse theories. If nonequilibrium violations of quantum theory, as I
envisage them, are \textit{not} observed, say over the next 100 years, then
I would start to have serious doubts about hidden variables (were I still
alive). I don't think the equilibrium de Broglie-Bohm theory by itself is
scientifically satisfactory, even if logically it is a possibility, because
the details of the trajectories can never be tested even in principle. I
might consider Everett's theory, if it can be shown to be conceptually
coherent.

I might also change my views, if someone made important progress in
theoretical physics -- for example about information loss in black holes, or
about quantum gravity -- for which a particular interpretation played a
crucial role.

\medskip \noindent \textbf{Q13. How do personal beliefs and values influence
one's choice of interpretation?} \\[-0.2cm]

It can be interesting and insightful to ask where people's ideas and
preferences come from, but only up to a point. In the end, what counts is
how the ideas stand up to theoretical and experimental scrutiny.

What I find more interesting is why you ask this question. It's a peculiar
question to ask a scientist. The whole point of science is to arrive at
objective truth, by a combination of reason and experiment, and to leave
personal beliefs by the wayside. Not to say that it's easy. But imagine
asking a biologist how `personal beliefs and values' affect the
interpretation of fossils. Or a condensed matter physicist, regarding the
interpretation of superconductivity. The question would seem peculiar, and
an insinuation that the person being questioned was or might be behaving
unscientifically by allowing personal beliefs to cloud their judgement.

Dennis Sciama used to say that, when it comes to the interpretation of
quantum mechanics, `the standard of argument suddenly drops to zero'. It's
still a field that is often short on argument and long on prejudice. It's as
if the usual rules of rational, scientific argument tend to be suspended in
this area. What have `personal beliefs and values' got to do with a
scientific discussion?

It might be claimed that the question is reasonable in the context of
quantum foundations, where there are different and radically divergent
interpretations. But our present uncertainty is no reason for compromising
basic standards. We should focus on arguments and evidence, not on beliefs.
The whole point of science is to get away from mere belief.

Some will say that realism in physics amounts to a `belief' or `value', but
that is confused and mistaken.

It is true that personal beliefs, values, and inclinations can provide
motivation and inspiration for new ideas or for following a certain road. In
principle, there's nothing wrong with that. Perhaps it's even necessary, if
one is going to make a serious effort exploring an idea. But again, once
ideas or research directions have been thought of and decided on, one has to
find out if there is any theoretical or experimental support for them. I
don't think it's scientifically healthy to give much weight to why someone
thought of something or why they're attracted to exploring a certain idea,
as it distracts from the scientific heart of the matter -- which is whether
or not the idea itself makes sense and is correct.

I would say that the widespread neglect of objectivity and realism in
quantum physics has contributed to an erosion of the idea of science as
finding out what the world is really made of and how it works. The objective
world is oblivious to our personal beliefs. Doubts about the existence of
the former tend to foster an emphasis on the importance of the latter.
Actually, the roots of this go back to the late eighteenth and early
nineteenth centuries, which saw the rise of German idealism as a force in
philosophy. The development of quantum mechanics itself, in the 1920s, shows
traces of its influence, with an emphasis on the subjective knowledge of the
human observer as opposed to objective reality.

Some people are uncomfortable with the objective and rational world of
science, which is sometimes seen as an impersonal world devoid of human
meaning, with no room for religious or spiritual belief. Sometimes, this is
what lies behind an attraction to subjective, non-ontological
interpretations of quantum physics. This seems to have been the case for
some of the founding quantum physicists, certainly in the case of Wolfgang
Pauli. But the trouble with subjective, non-ontological interpretations is
not so much the motivation that sometimes lies behind them, but the fact
that they don't make sense (see my discussion of the measurement problem in
my answer to Question 7).

\medskip \noindent \textbf{Q14. What is the role of philosophy in advancing
our understanding of the foundations of quantum mechanics?}

I would say that our thinking about quantum physics became muddled in the
1920s, under the influence of certain incorrect philosophical ideas that
were fashionable in some circles at the time. I think physicists need to
un-learn some of those wrong ideas, in order to return to clear scientific
thinking about quantum theory. In this, some exposure to analytical
philosophy can be helpful. For example, any graduate student in the subject
knows that, if anything, indeterminism makes free will even harder to
explain, since our actions would be occurring for no reason, and yet one
often hears physicists citing our apparent free will as a reason for
abandoning determinism.

On the other hand, if people in quantum foundations would start thinking
like other physicists and scientists do -- in terms of an objective reality
that we need to discover -- there would be little need for philosophy. For
example, spectroscopic analysis enables us to deduce the chemical
composition of the stars, despite Auguste Comte's infamous claim in the
nineteenth century that this would never be possible. Yet, astrophysicists
don't need to study philosophy. Similarly, biologists and geologists have
deduced that certain events occurred on earth millions of years ago, without
worrying about philosophical questions.

I've already mentioned, in my answer to Question 13, the unfortunate effects
of German idealism on scientific thought. That philosophical movement was
deeply influential in Denmark, as well as in the German-speaking world. It
was sparked off by the publication in 1781 of Kant's \textit{Critique of
Pure Reason}, a work that was widely interpreted as having undone the
Copernican revolution and having restored human beings to a central place in
the order of things, a claim that was later repeated in 1927 with the rise
of quantum mechanics. Kant, like many others of his time, believed that
Newton had discovered the true physics. On the other hand, David Hume argued
that certain knowledge about the world was impossible to acquire. Kant was
faced with a paradox. How had Newton done it? Kant's ingenious answer, as
least as widely interpreted, was that Newtonian physics reflected the
structure of human thought, not the structure of the world, and that the
world `in itself' was unknowable. Now, a lot happened between 1781 and 1927.
But if we ask why Bohr and Heisenberg took seriously the absurd idea that
experiments \textit{must} be described in terms of classical physics -- a
claim that is easily refuted by describing experiments in terms of, for
example, de Broglie's nonclassical pilot-wave dynamics of 1927 -- then the
answer is that they took seriously the bizarre Kantian claim that classical
physics is essential to the structure of human thought. In order to clear
away such wrong-headed ideas, it helps to know where they came from and why
they were proposed. For this purpose, a knowledge of the history of
philosophy can be helpful.

Another wrong idea that needs clearing away is operationalism. This has
roots in the philosophy of Ernst Mach, and entered physics with the
publication of Einstein's first relativity paper in 1905. For the first
time, the human observer seemed to play a central role in physics. More
substantially, as I've already said in my answer to Question 10, Einstein
made the fatal mistake of treating macroscopic equipment -- in Einstein's
case, rods and clocks -- as if they were fundamental objects, when in fact
they are emergent and approximate objects built out of more elementary
things such as particles and fields. The widespread respect for Einstein's
approach seemed to justify Bohr's subsequent belief that classical apparatus
played a fundamental role in quantum theory, even though any apparatus is
built out of nonclassical atoms. Another mistake propagated by Einstein's
first paper, at least as it was widely interpreted, is the idea that physics
should be based only on what is observable, when in fact, as Einstein later
explained to Heisenberg in a private conversation in 1926, some body of
theory is required before we can even know how to make reliable observations
-- what we can observe is determined by the theory, not the other way round.
To correct these kinds of mistakes, analytical philosophy can be helpful.

But at the end of the day, what is needed, essentially, is clear and
objective scientific thinking, of the sort exemplified by Bell -- what was
once normal practice, among people like Maxwell, Boltzmann, and de Broglie,
for example, and what to this day remains normal practice in almost every
branch of science, including physics. The essential point, shared by almost
all scientists, is that there is a real world, and that it is the task of
science to find out about it. In the narrow context of quantum foundations I
would say that, for the most part, we need to \textit{un}-learn some bad
philosophical ideas that have become associated with the subject, and which
scientists in other areas would never take seriously.

\medskip \noindent \textbf{Q15. What new input and perspectives for the
foundations of quantum mechanics may come from the interplay between quantum
theory and gravity/relativity, and from the search for a unified theory?} \\[%
-0.2cm]

There has, of course, already been a lot of input. Work in quantum gravity
often has a cosmological setting, where in the very early universe the lack
of an external classical background makes textbook quantum theory
inadequate. This was, historically, one of the motivations for the Everett
interpretation. Today, according to inflationary cosmology, the remnants of
primordial quantum fluctuations are imprinted on the cosmic microwave
background, and a proper understanding of the quantum-to-classical
transition during the inflationary era again forces us to think beyond the
textbooks. In the context of a theory like inflation, which is currently
being tested experimentally, the Copenhagen interpretation can't be taken
seriously. Quantum foundations needs to catch up with what has been going on
elsewhere in fundamental physics and cosmology. How can we, for example,
return to something like `operational quantum theory', which relies on a
classical background containing macroscopic apparatus, when there is an
experimental need to discuss a quantum-to-classical transition that took
place in the earliest moments after the big bang?

I'd also like to point out that there is currently a great opportunity to
use cosmology as a testing ground for quantum theory under new and extreme
conditions, at very short distances and very high energies. Inflationary
cosmology, in particular, is being used as a laboratory to test almost every
modification of high-energy physics that theorists are able to think of, and
yet hardly anyone is using it to test quantum theory itself. A handful of
people, such as Daniel Sudarsky, have considered how to use it to test
collapse theories, and I have studied how to use it to test for quantum
nonequilibrium in the early universe.\footnote{%
A. Valentini, `Inflationary cosmology as a probe of primordial quantum
mechanics', \textit{Phys. Rev. D} \textbf{82}, 063513 (2010).} But there is
a vast amount of further work that could and should be done.

On the subject of gravity proper, there is the puzzle of black hole
information loss, and the alarming possibility that a closed system can
evolve from a pure to a mixed state. This problem has fuelled an immense
amount of work in high-energy physics and string theory, where it is hoped
that ideas like the AdS/CFT correspondence will provide a fundamentally
unitary description of black hole formation and evaporation. This is an
important problem in quantum foundations, and I've speculated, in a
hidden-variables context, that Hawking radiation may consist of
nonequilibrium particles that violate the Born rule -- where such states can
carry more information than conventional quantum states. That's a line of
thought I hope to develop further.

The nonlocality of de Broglie-Bohm theory, and of hidden-variables theories
generally, points to the existence of an absolute time. This might help with
solving the notorious `problem of time' in quantum gravity. But already at
the level of standard quantum gauge theories, in Minkowski spacetime, there
is a tension with manifest Lorentz covariance, which requires the
introduction of bosonic `ghost' states with negative norm. I've always
thought that the simplicity of non-covariant and ghost-free gauges, such as
the temporal gauge, in theories such as quantum chromodynamics, points to
the existence of an underlying preferred state of rest, and I find the
pilot-wave version of gauge field theory -- at least as I formulate it, with
three-vector gauge fields on an Aristotelian spacetime -- to be more elegant
than the standard version. I suspect that this line of thought may be worth
developing further.

I find it interesting that the AdS/CFT correspondence might be interpreted
as saying that physics is really based on a Yang-Mills gauge theory on flat
spacetime. It would be straightforward to make a de Broglie-Bohm version of
the latter, and one has to wonder how the underlying preferred frame would
relate to the emergent gravitational description.

Finally, it wouldn't be surprising if one particular interpretation of
quantum theory proved to be crucial in developing a unified theory. But we
won't find out for as long as quantum foundations remains so removed from
the rest of physics.

\medskip \noindent \textbf{Q16. Where would you put your money when it comes
to predicting the next major development in the foundations of quantum
mechanics?} \\[-0.2cm]

I suspect that the field of quantum foundations will develop properly only
when it starts attracting people who are fully conversant with modern
theoretical physics and its problems at an advanced level. There has been
too much work on elementary quantum mechanics, for example for simple
entangled systems. As I said in my reply to Question 15, there are important
theoretical problems concerning the early universe, black holes, and quantum
gravity, and exploring these further might lead to something important. But
if I was asked to make one guess, then if the history of physics is anything
to go by, I'd say it's most likely that an experimental breakthrough will be
needed to make further progress.

Some areas of theoretical physics have lost contact with physics as an
experimental science, not only in the trivial sense that we don't have
experimental anomalies that defy explanation with current theories, but also
in the more sinister sense that some theorists grossly underestimate the
crucial role that experiment has played in the historical development of our
theories. Contary to folklore, no really major advance in fundamental
physics has ever occurred without guidance from experiment. Theorists like
to believe that pure thought can suffice, and they often cite the example of
general relativity in 1915, where textbooks and popular accounts often give
the impression that the correct perihelion motion of Mercury came out of the
theory as an unexpected bonus. The truth is, Einstein used the observed
perihelion motion to rule out his own pre-1915 theory of gravity. Once he
found the field equations that gave the correct perihelion motion, only then
could he be confident in the other predictions coming from those same
equations. And there are plenty of other examples. Schr\"{o}dinger's
original wave equation was beautiful and Lorentz invariant -- and it gave
the wrong energy levels for hydrogen. His non-relativistic version seemed
less elegant, but gave the right energy levels. And so on. If we look at the
examples that theorists often cite, and if we examine what really happened
historically as opposed to the folklore in theoretical textbooks, we always
find that experiment played a much bigger role than theorists like to
believe.

So it's likely that we need an experimental clue, an empirical window. To
that end, we should be trying harder to test quantum mechanics in genuinely
new and extreme domains. My `prediction', for what it's worth, is that we
will find an experimental breakdown of quantum theory. My guess is that
quantum theory will turn out to be an equilibrium case of a broader theory
based on hidden variables, and this motivates me to suggest experiments
searching for nonequilibrium violations of quantum theory.

\medskip \noindent \textbf{Q17. What single question about the foundations
of quantum mechanics would you put to an omniscient being?} \\[-0.2cm]

My question would be: `Is it possible in principle to use entangled systems
for superluminal signalling?'

If the answer was `yes', I would hope that I was on the right track (though
nonlinearities might yield similar effects). If the answer was `no', I would
abandon deterministic hidden-variables theories.

\end{document}